\documentclass[twocolumn,aps,superscriptaddress]{revtex4-1}
\usepackage{amsfonts}
\usepackage{amsmath}
\usepackage{amssymb}
\usepackage{graphicx}
\usepackage{color}
\usepackage{ulem}
\usepackage{mathrsfs}
\usepackage[colorlinks, urlcolor=cyan, citecolor=blue, linkcolor=magenta]{hyperref}

\begin{document}

\title{Fate of the spatial-temporal order under quantum fluctuation}
\author{Xiaotian Nie}
\affiliation{Hefei National Laboratory, University of Science and Technology of China,
Hefei 230088, China}
\author{Wei Zheng}
\email{zw8796@ustc.edu.cn}
\affiliation{Hefei National Laboratory, University of Science and Technology of China,
Hefei 230088, China}
\affiliation{Hefei National Laboratory for Physical Sciences at the Microscale and Department of Modern Physics, University of Science and Technology of China,
Hefei 230026, China}
\affiliation{CAS Center for Excellence in Quantum Information and Quantum Physics,
University of Science and Technology of China, Hefei 230026, China}
\date{\today }

\begin{abstract}
	In a previous theoretical work [arXiv:2205.01461], T. Esslinger group proposed a scheme to realize a spatial-temporal lattice, which possesses dual periodicity on space and time, in a cavity-boson system pumped by a travelling wave laser. However, the prediction was made under the mean-field approximation. In this work, we investigate the dynamics beyond mean-field approximation. By including the fluctuation of the cavity field, we obtain a larger set of equations of motion. Numerical results show that the spatial-temporal lattice is melted in the mean-field level but survives in the quantum fluctuation. 
\end{abstract}

\maketitle

\section{Introduction}

Optical cavities coupled with ultracold atoms provide powerful tools to simulate non-equilibrium many-body quantum systems.  Inevitable photon leaking results in the dissipation and drives the system away from equilibrium. In such cavity-atom hybridized systems, most researches are aiming at two main directions. The first one is to deal with various steady states and their transitions, like superradiance, multi-stability and hysteresis\cite{Cavity-Dicke@Esslinger.2010,ExpBsSr@Hemmerich.2015,Exp1stNondiss@Esslinger.2021,ExpFmSr@WHB.2021,ThyBsSr@Jia.2020,ThyBsSr@Ritsch.2002,ThyBsSr@Ciuti.2014,ThyBsSr@Zilberberg.2018,DoubleSymmetryBreaking@nataf.2012,ThyBsUns@Keeling.2012,KeldyshDicke@Diehl.2013,QFT-OS@Diehl.2016,Lieb.1973,Dicke@Hioe.1973,ThyBsSr@Carmichael.2007,ThyBsSr@Domokos.2008,ThyBsSr@Vukics.2005,ThyBsSr@Irish.2017,ThyFmSr@Ritsch.2019,ThyFmSr@Kollath.2016,ThyFmSr@Keeling.2014,ThyFmSr@CY.2014,ThyFmSr@CY.2015,ThyFmSr@Piazza.2017,ThyFmSr@YW.2018,ThyFmSr@Brennecke.2016,ThyFmSr@Piazza.2014,NonequilibriumPhasesFermi@nie.2023,FluctuationinducedBistabilityFermionic@tolle.2024,ManyBodyOpenQuantum@fazio.2024,ExceptionalPointHysteresis@zhang.2024}.

The other direction is to focus on the dynamically unstable phases which can never reach a steady state\cite{ExpBsUns2@Esslinger.2019,ExpBsUns@Esslinger.2022}. Some of them are just chaotic without regularity\cite{ExpBsUns@Esslinger.2019,NonequilibriumPhasesFermi@nie.2023,DynamicalPhasesBEC@harmon.2024}. A specific kind of unstable phase has been attracting more and more interest in the past decade. Time crystals, which spontaneously break the time translation symmetry and show permanent periodic oscillation, have been predicted and observed in such systems\cite{OTC@Piazza.2015,DickeCTC@Keeling.2018,CTC2@Jaksch.2019,ThyBsUnst@Nunnenkamp.2019,OTC@Tuquero.2022,ThyBsUns@Keeling.2012,CTC@Hemmerich.2019,CTC@XWL,Zheng.2023,CTC@Hemmerich.2022,TorusBifurcationDissipative@cosme.2024,ManyBodyOpenQuantum@fazio.2024,QuantumOriginLimit@dutta.2024,ExperimentalRealizationDiscrete@he.2024a,TopologicalTransportClassical@simula.2024,ObservationTimeCrystal@jiao.2024}.

Apart from the time crystals, more and more novel non-equilibrium dynamically unstable phases are predicted and observed\cite{ExperimentalRealizationDiscrete@he.2024}. T. Esslinger group proposed a theoretical scheme to realize a spatial-temporal lattice, which possesses double periodicities on space and time\cite{ThyBsUns@Esslinger.2022}. That is to say, the amplitude of the self-organized lattice oscillates periodically. It should be emphasized that this spatial-temporal lattice does not belong to time crystals, since no matter the phase of photon field or the atomic momentum can not return to the original state after a period. Although this phenomenon is quite interesting, it is predicted under the mean-field approximation. Whether it will survive in or be destroyed by quantum fluctuation is an open question. 

In this paper, we take a step further to include the effect of quantum fluctuation, and investigate how those dynamically unstable phases will be influenced. By separating the mean-field part and the fluctuation, new equations of motion are obtained. Numerical results indicate that the mean value of the lattice vanishes. However, the fluctuation of the lattice will carry on oscillating with a shorter period, which means the spatial-temporal lattice phase survives only in quantum fluctuation level.
Besides, we also investigate the other two non-equilibrium phases predicted by the mean-field approximation, their behaviors are not changed qualitatively.

\section{The setup and model}

\begin{figure}[!h]
    \centering
    \includegraphics[width=0.5\textwidth]{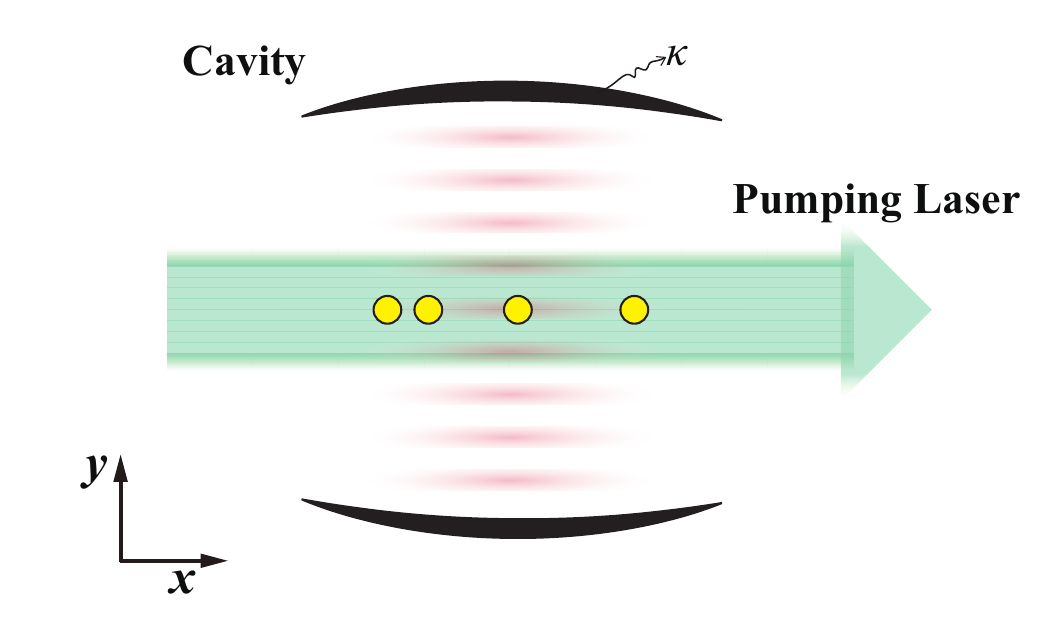}
    \caption{Schematic illustration of the physical setup. One-dimensional bosonic atoms are coupled to a transverse optical cavity illuminated by one single pumping beam. The atoms are confined along the pumping laser.}
    \label{setup}
\end{figure}

Fig.\ref{setup} shows the physical setup, where one-dimensional bosonic atoms restricted in the $x$-axis are loaded into an optical cavity set along the $y$-axis. The 
atoms are illuminated by a beam of travelling wave laser forward the $+x$-direction. In such a setup, the atoms can only feel an interference lattice, which describes the second-order process of the atoms scattering a photon from the pumping laser to the cavity mode or vice versa. In the momentum space, the Hamiltonian of this cavity-atom system is given by\cite{ThyBsUns@Esslinger.2022}
\begin{eqnarray}
	H &=& -\Delta \hat{a}^\dagger \hat{a} + \sum_i \xi_i \hat{c}_i^\dagger \hat{c}_i\nonumber\\
	&&+\eta\sum_i\left(\hat{a}^\dagger \hat{c}^\dagger_{i+1} \hat{c}_{i}+a \hat{c}^\dagger_{i} \hat{c}_{i+1}\right),
\end{eqnarray}
where $\Delta$ is detuning between the cavity mode and the pumping frequency, $a$ is the annihilation operator of the cavity photon with momentum $k_\mathrm{ph}$, $\hat{c}_i$ is the bosonic annihilation operator of the atom at momentum $k_i=i\times k_\mathrm{ph}$, $\xi_i = \frac{k_i^2}{2m}$ is the corresponding kinetic energy and $\eta$ is the interaction strength proportional to the pumping amplitude. Due to the photon leaking of the cavity, the evolution of the system is governed by the Lindblad master equation,
\begin{eqnarray}
	\partial_t \rho(t)=-\mathrm{i}[H,\rho]+\kappa\left(2\hat{a}\rho \hat{a}^\dagger-\hat{a}^\dagger \hat{a} \rho-\rho \hat{a}^\dagger \hat{a}\right).
\end{eqnarray}
The equations of motion of the photonic and atomic operators are
\begin{eqnarray}
	\mathrm{i}\partial_t \hat{a}&=&(-\Delta-\mathrm{i}\kappa)\hat{a}+\eta\sum_l \hat{c}^\dagger_{l+1} \hat{c}_l,\\
	\mathrm{i}\partial_t \hat{c}_i&=& \xi_i \hat{c}_i+\eta\left(\hat{a}^\dagger \hat{c}_{i-1}+\hat{a} \hat{c}_{i+1}\right).
\end{eqnarray}

\section{Mean-field Analysis}

This model is qualitatively distinct from previous similar ones, such as cavity-atom systems pumped by a laser pair. In those models, the system can hold several steady phases by simultaneously scattering photons from two opposite directions. However, in this travelling wave model, only photons with $+k_\mathrm{ph}$ are injected into the system. So atoms will keep acquiring momentum from the pumping laser, leading to unstable dynamics.

The non-equilibrium phases are distinguished by their dynamical evolution. To solve that, in the original proposal\cite{ThyBsUns@Esslinger.2022}, the mean-field approximation is adopted by assuming the cavity is always in a coherent state, and the entanglement between the atoms and the cavity is negligible. Then, the evolution is governed by

\begin{eqnarray}
	\mathrm{i}\partial_t \alpha&=&(-\Delta-\mathrm{i}\kappa)\alpha+\eta\sum_l \rho_{l+1,l},\\
	\mathrm{i}\partial_t \rho_{ij}&=&(\xi_j-\xi_i)\rho_{ij}+\eta\alpha^*\rho_{i,j-1}+\eta\alpha\rho_{i,j+1}\nonumber\\
	&&-\eta\alpha^*\rho_{i+1,j}-\eta\alpha\rho_{i-1,j},\label{MFrho}
\end{eqnarray}
where $\alpha(t)=\left\langle \hat{a} (t)\right\rangle$ is the mean field of cavity, and the single-particle density matrix of the atoms is $\rho_{ij}=\left\langle \hat{c}_i^\dagger \hat{c}_j\right\rangle$.

There, they calculate the evolution under different parameters and find three distinct non-equilibrium dynamical phases: the spatial lattice, the spatial-temporal lattice and the dephasing phase. In the spatial lattice, the interference lattice depth roughly maintains constant (though the phase of the lattice keeps increasing, leading to the atoms continuously accelerating). While the spatial-temporal lattice is a highlighted new dynamical phenomenon due to its double periodicity, where the lattice appears and vanishes periodically over time. In the dephasing phase, the evolution is chaotic without any regular features. (All these mean-field results are also shown later in our figures to compare with our results.)

In their paper, they explain the emergence of the spatial-temporal lattice as a consequence of a self-organized $\pi$-pulse between atomic neighboring momentum states induced by the cavity field\cite{ThyBsUns@Esslinger.2022}. This self-organized $\pi$-pulse interpretation leaves a concern to readers that the model may need fine-tune, any perturbation could destroy this phenomenon. Whether this double-periodicity can survive under quantum fluctuation is an open question\cite{ZW@2016}.

\section{Beyond mean field}


In this paper, we take a step further to include the effect of quantum fluctuation. To distinguish the mean-field part and quantum fluctuation, we are going to separate the cavity operator to its expectation value and the fluctuation, $\hat{a}(t)=\alpha(t)+\delta \hat{a} (t)$, where the fluctuation $\delta \hat{a} (t)$ is still an operator satisfying bosonic commutation relation, $[\delta \hat{a}, \delta \hat{a}^\dagger]=1$. And by definition, $\delta \hat{a} (t)$ has vanishing expectation value.

So, the equations of motion are revised to:
\begin{eqnarray}
	\mathrm{i}\partial_t \alpha&=&(-\Delta-\mathrm{i}\kappa)\alpha+\eta\sum_l \rho_{l+1,l},\\
	\mathrm{i}\partial_t \delta \hat{a}&=&(-\Delta-\mathrm{i}\kappa)\delta \hat{a}+\eta\sum_l \left\{ c^\dagger_{l+1} \hat{c}_l-\rho_{l+1,l}\right\},\\
	\mathrm{i}\partial_t \hat{c}_i&=& \xi_i \hat{c}_i+\eta \left(\alpha^\ast \hat{c}_{i-1}+\alpha \hat{c}_{i+1}\right.\nonumber\\
	&&\left.\qquad \qquad+\delta \hat{a}^\dagger \hat{c}_{i-1}+\delta \hat{a} \hat{c}_{i+1}\right) .
\end{eqnarray}
Then, the time evolution of $\rho_{ij}$ becomes
\begin{eqnarray}
	\label{EoM of DM}
	\mathrm{i}\partial_t \rho_{ij} &=& (\xi_j-\xi_i)\rho_{ij}\\
	&&+\eta\alpha^*\rho_{i,j-1}
	+\eta\alpha\rho_{i,j+1}\nonumber\\
	&&-\eta\alpha^*\rho_{i+1,j}
	-\eta\alpha\rho_{i-1,j}\nonumber\\
	&&+\eta\left\langle \delta \hat{a}^\dagger \hat{c}_i^\dagger \hat{c}_{j-1}\right\rangle
	+\eta\left\langle \delta \hat{a} \hat{c}_i^\dagger \hat{c}_{j+1}\right\rangle\nonumber\\
	&&-\eta\left\langle \delta \hat{a}^\dagger \hat{c}_{i+1}^\dagger \hat{c}_{j}\right\rangle
	-\eta\left\langle \delta \hat{a} \hat{c}_{i-1}^\dagger \hat{c}_{j}\right\rangle\nonumber.
\end{eqnarray}
The upper equations are not closed yet, so we should further consider the equation of $\left\langle \delta \hat{a} \hat{c}_{i}^\dagger \hat{c}_{j}\right\rangle$, it is
\begin{eqnarray}
	\label{EoM of 3}
	\mathrm{i}\partial_t \left\langle \delta \hat{a} \hat{c}_{i-1}^\dagger \hat{c}_{j}\right\rangle &=& (\xi_j-\xi_i-\Delta-\mathrm{i}\kappa)\left\langle \delta \hat{a} \hat{c}_{i-1}^\dagger \hat{c}_{j}\right\rangle\\
	&&+\eta\alpha^*\left\langle \delta \hat{a} \hat{c}_{i}^\dagger \hat{c}_{j-1}\right\rangle
	+\eta\alpha\left\langle \delta \hat{a} \hat{c}_{i}^\dagger \hat{c}_{j+1}\right\rangle\nonumber\\
	&&-\eta\alpha^*\left\langle \delta \hat{a} \hat{c}_{i+1}^\dagger \hat{c}_{j}\right\rangle
	-\eta\alpha\left\langle \delta \hat{a} \hat{c}_{i-1}^\dagger \hat{c}_{j}\right\rangle\nonumber\\
	&&+\eta\left\langle \delta \hat{a}\delta \hat{a}^\dagger \hat{c}_{i}^\dagger \hat{c}_{j-1}\right\rangle
	+\eta\left\langle \delta \hat{a}\delta \hat{a} \hat{c}_{i-1}^\dagger \hat{c}_{j+1}\right\rangle\nonumber\\
	&&-\eta\left\langle \delta \hat{a}\delta \hat{a}^\dagger \hat{c}_{i+1}^\dagger \hat{c}_{j}\right\rangle
	-\eta\left\langle \delta \hat{a}\delta \hat{a} \hat{c}_{i-1}^\dagger \hat{c}_{j}\right\rangle\nonumber\\
	&&+\eta\sum_l\left\{ \left\langle \hat{c}_{l+1}^\dagger \hat{c}_l \hat{c}_i^\dagger \hat{c}_j\right\rangle- \rho_{l+1,l}\rho_{ij}\right\}.\nonumber
\end{eqnarray}
We can decouple the quartic terms to quadratic ones by cumulant expansion\cite{DickeCTC@Keeling.2018,NoiseresilientPhaseTransitions@alaeian.2024,SymmetriesCorrelationsContinous@mukherjee.2024},
\begin{eqnarray}
	\left\langle \delta \hat{a}\delta \hat{a}^\dagger \hat{c}_{i+1}^\dagger \hat{c}_{j}\right\rangle &\approx&
	\left\langle \delta \hat{a}\delta \hat{a}^\dagger \right\rangle \rho_{i+1,j}\nonumber\\
	\left\langle \hat{c}_{l+1}^\dagger \hat{c}_l \hat{c}_i^\dagger \hat{c}_j\right\rangle &\approx& \rho_{l+1,l}\rho_{ij} + \left(\rho_{li}+\delta_{li}\right)\rho_{l+1,j}\nonumber.
\end{eqnarray}
Approximately, we ignore the atomic kinetic energy and the four terms proportional to $\alpha$ (or $\alpha^*$) in Eq.\ref{EoM of 3} by comparing them with the first term
, and then take the pairing term of fluctuation $\left\langle \delta \hat{a}\delta \hat{a} \right\rangle \approx 0$. Now following the process of adiabatic elimination, we assume $\mathrm{i}\partial_t \left\langle \delta \hat{a} \hat{c}_{i}^\dagger \hat{c}_{j}\right\rangle=0$, obtaining
\begin{eqnarray}
	\label{Adiabatic Elimination}
	\left\langle \delta \hat{a} \hat{c}_{i}^\dagger \hat{c}_{j}\right\rangle&=&\frac{\eta}{\Delta+\mathrm{i}\kappa}\left\{(\delta n +1)\rho_{i,j-1}\right.\\
	&&-\delta n \rho_{i+1,j}\left.+\sum_l \rho_{il}\rho_{l+1,j}\right\}\nonumber,
\end{eqnarray}
where we denote $\delta n = \left\langle \delta \hat{a}^\dagger \delta \hat{a}\right\rangle $, which means the extra photon number due to fluctuation, total photon number should be $n=|\alpha|^2+\delta n$. Further, the equation of motion of $\delta n$ is
\begin{eqnarray}
	\mathrm{i}\partial_t \delta n &=&   -2\mathrm{i}\kappa \delta n +	\eta \sum_l \left\{\left\langle\delta \hat{a}^\dagger \hat{c}_{l+1}^\dagger \hat{c}_{l}\right\rangle-\left\langle\delta \hat{a} \hat{c}_{l}^\dagger \hat{c}_{l+1}\right\rangle\right\}\nonumber\\
	&=&-2\mathrm{i}\kappa\delta n + \frac{2\mathrm{i}\kappa\eta^2}{\Delta^2+\kappa^2}\left(N+\sum_k\sum_l \rho_{kl}\rho_{l+1,k+1}\right)\nonumber,
\end{eqnarray}
where $N$ is the total atom number.
In the last equation, we use the approximation $\xi_i\ll |\Delta|,\kappa$, as the atomic recoil energy is far less than the cavity energy scale in practice.
Substituting Eq.\ref{Adiabatic Elimination} into Eq.\ref{EoM of DM}, then the equation of motion of density matrix is (sum over $l$ is omitted for short):
\begin{eqnarray}
	&&\partial_t \rho_{ij}(t)+\mathrm{i}\left[h_{jl}(t)\rho_{il}-\rho_{lj}h_{li}(t)\right]\\
    &&=\frac{\kappa\eta^2}{\Delta^2+\kappa^2}\times\nonumber\\
	&&\left\{2(\delta n +1)(\rho_{i-1,j-1}-\rho_{i,j})-2\delta n(\rho_{i,j}-\rho_{i+1,j+1})\right.\nonumber\\
    &&\left.+ \rho_{i-1,l-1}\rho_{lj}+  \rho_{il}\rho_{l-1,j-1}- \rho_{i+1,l+1}\rho_{lj}- \rho_{il}\rho_{l+1,j+1}    \right\}\nonumber\\
	&&\quad+\frac{\mathrm{i}\Delta\eta^2}{\Delta^2+\kappa^2}\times\nonumber\\
	&&\left\{ \rho_{i+1,l+1}\rho_{lj}+\rho_{i-1,l-1}\rho_{lj} - \rho_{il}\rho_{l-1,j-1}-\rho_{il}\rho_{l+1,j+1}\right\}
	,\nonumber
\end{eqnarray}
where $h_{ij}=\xi_i \delta_{ij}+\eta \alpha^*\delta_{i-1,j}+\eta\alpha\delta_{i+1,j}$ is the mean-field single-atom Hamiltonian matrix in the momentum space.

If the right side of the equation above is equal to zero, it is nothing but the mean-field equation of motion (see Eq.\ref{MFrho}). The right side represents purely quantum fluctuation of atoms. We could see that its second part is just the self-energy correction that can be ignored, while the first part is dominant as the fluctuation-induced current\cite{ZW@2016}. The prefactor ${\kappa}/{\Delta^2+\kappa^2}$ suggests the fluctuation of atoms will be suppressed when the photonic dissipation is either too weak or too strong. It is because that due to the fluctuation-dissipation relation, the photonic fluctuation is proportional to the photonic dissipation\cite{FieldTheoryNonequilibrium@kamenev.2011,QuantumFieldTheory@rammer.2011,KeldyshDicke@Diehl.2013,QFT-OS@Diehl.2016}. Consequently, the photonic fluctuation will drive the atoms to fluctuate. In the weak dissipation regime, they are positively correlated. But when the dissipation becomes too large, the photon field stays near the vacuum, coupling between the atoms and cavity is weak, and the fluctuation of atoms will also be suppressed.

This result is consistent with the one in a similar model calculated by Keldysh formalism \cite{ZW@2016} instead of equations of motion here. Moreover, our method also contains the quantum fluctuation of cavity field $\delta n$.

\section{Numerical simulation}

\begin{figure}[!b]
    \centering
    \includegraphics[width=0.5\textwidth]{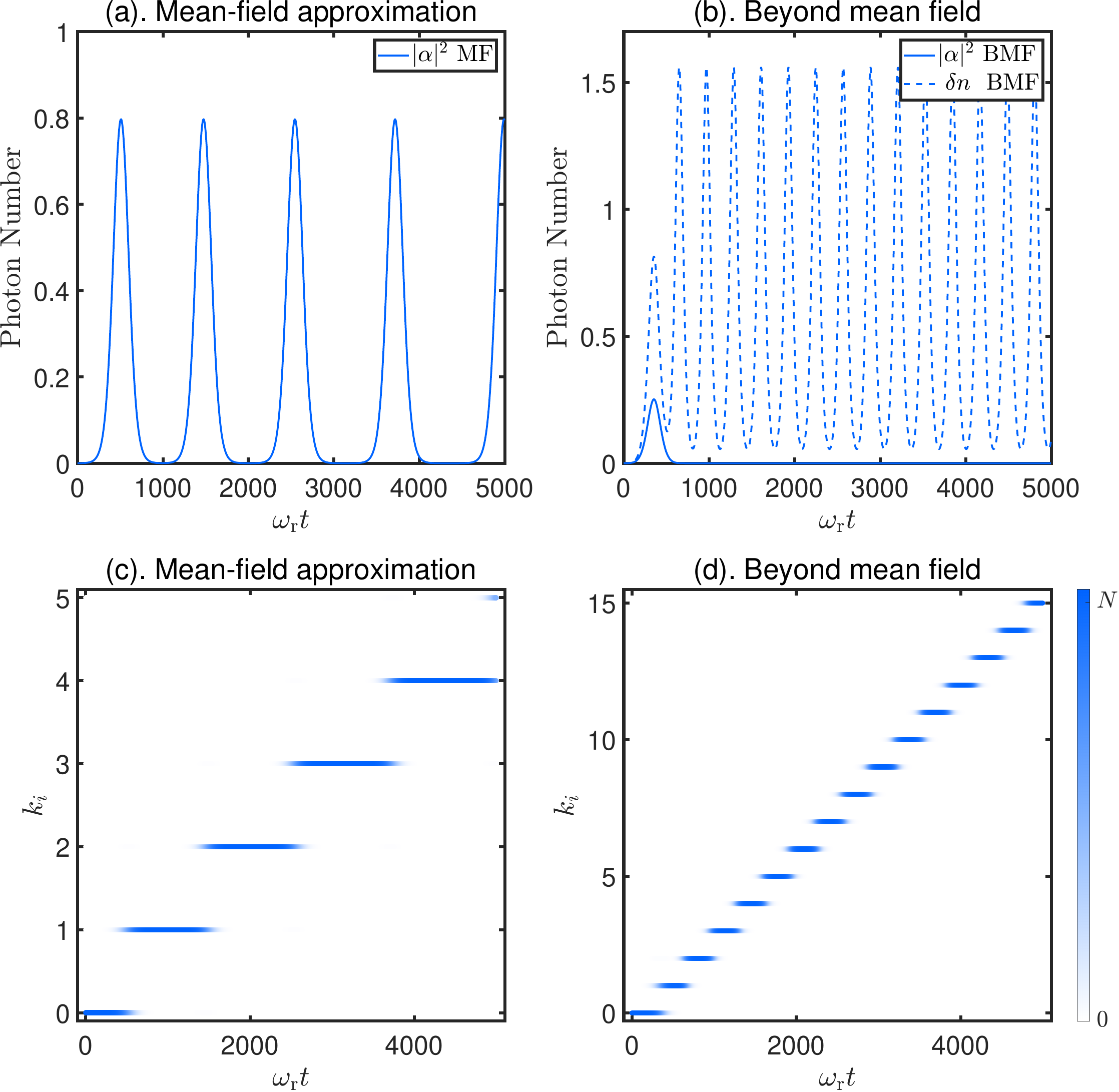}
    \caption{Evolution of the spatial-temporal lattice, the parameters are $\Delta = 2\pi\times 40~\mathrm{MHz}$, $\eta=2\pi\times 1~\mathrm{kHz}$. (a). The $|\alpha|^2$ oscillation in the mean-field approximation\cite{ThyBsUns@Esslinger.2022}. (b). Two parts of the photon occupation beyond mean field (BMF), $|\alpha|^2$ and $\delta n$. After one single oscillation, mean-field part vanishes and the fluctuation keeps oscillating. The period is shorter than the mean-field result. (c).\&(d). The atomic momentum distribution over time within and beyond the mean-field approximation. The transfer rate is enhanced by fluctuation.}
    \label{STL}
\end{figure}


By numerically solving the equations, we obtain the dynamical evolution of the expectation value of photon field, fluctuation of photon number and the atomic momentum distribution. For the convenience of experimental reference, we set the parameters practically, atomic recoil energy (for $^{87}\mathrm{Rb}$) $\omega_\mathrm{r}=k_{\mathrm{ph}}^2/2m=2\pi\times 3.7~\mathrm{kHz}$, cavity linewidth $\kappa=2\pi\times 1~\mathrm{MHz}$ and total atom number $N=1\times 10^5$\cite{ThyBsUns@Esslinger.2022}. Following the original proposal, we set the initial state as a direct product of the photonic vacuum and a macroscopic atomic occupation on a perturbative superposition of $k_0$ and $k_1$: $\sqrt{\mathcal{N}^{-1}}(\hat{c}_0^\dagger+\epsilon \hat{c}_1^\dagger)^N |0\rangle_\mathrm{atom}$, where $\mathcal{N}$ is a normalization factor and $\epsilon=0.01$.

Here, we mainly focus on the behavior of the spatial-temporal lattice. As shown in Fig.\ref{STL}, $|\alpha|$ remains nearly zero after only one single oscillation, which means the expectation value of the interference lattice depth vanishes quickly. But it does not imply the "spatial-temporal lattice" totally melts or the atoms will stay still forever. Due to the quantum fluctuation of the photon field, the total photon number still oscillates with a shorter period. And the atoms are yet pumped to higher momentum states. That means the "spatial-temporal lattice" survives under the quantum fluctuation in another form.

Although the value of the temporal period seems to be strongly influenced by the fluctuation, the periodicity still exists and shows robustness. That is because the spatial-temporal lattice exists in the weak pumping regime\cite{ThyBsUns@Esslinger.2022}, only two neighboring momentum states are macroscopically occupied simultaneously. So in a specific atomic momentum transfer cycle, we can consider only these two momentum states (i.e. in the cycle: $i\to i+1$, $\rho_{m,n\notin\{i,i+1\}}=0$). 
The only difference between this cycle: $i\to i+1$ and the next cycle: $i+1\to i+2$ process is the atomic kinetic energy difference $\xi_{i+1}-\xi_i$ and $\xi_{i+2}-\xi_{i+1}$. But this difference can be totally removed in a rotating frame. Therefore, every cycle costs the same time, leading to the temporal periodicity.

\begin{figure}[!h]
    \centering
    \includegraphics[width=0.48\textwidth]{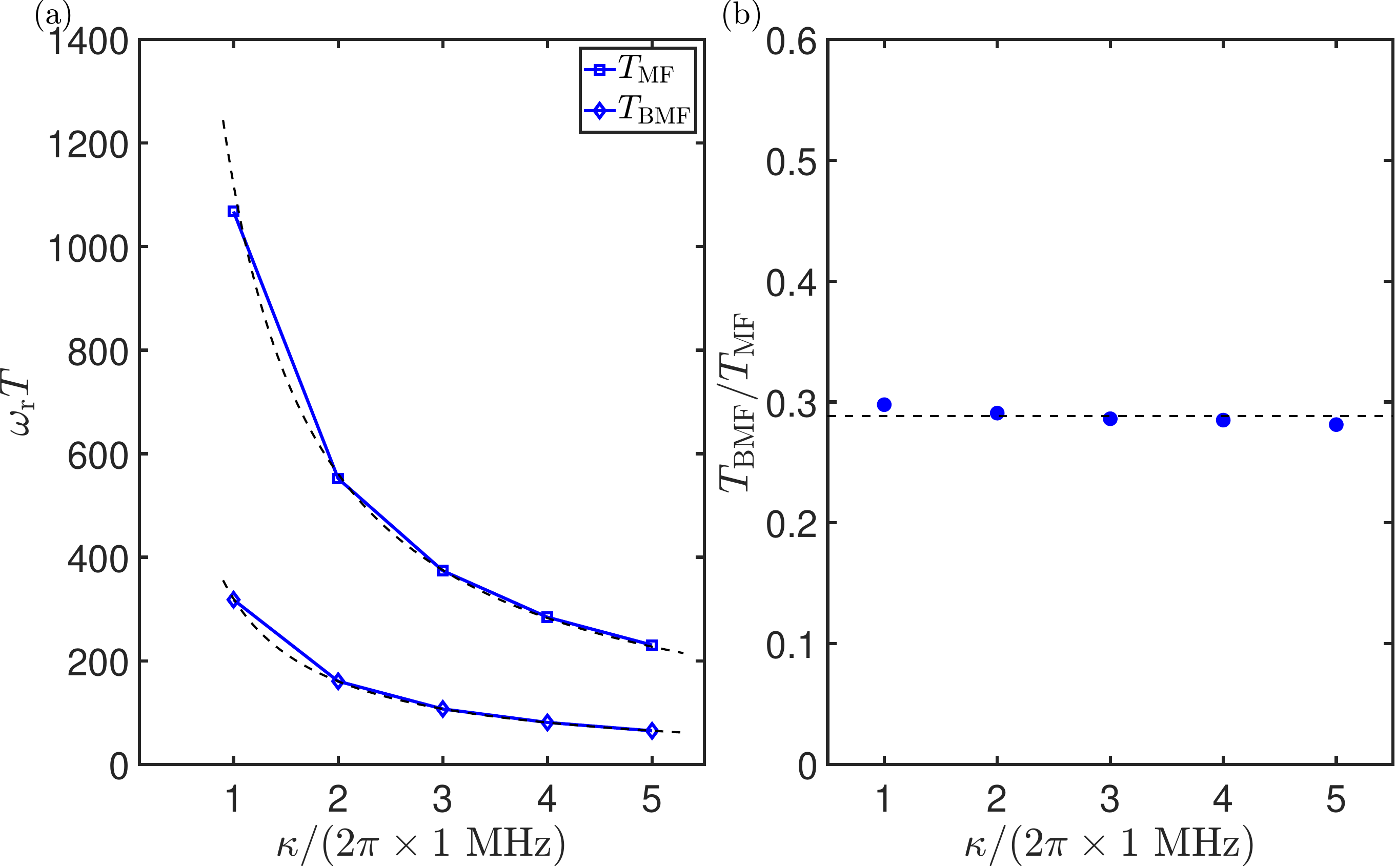}
    \caption{(a) Oscillation period of our result ($T_\mathrm{BMF}$) and the original mean-field result ($T_\mathrm{MF}$). They both fit well with $T^{-1}\propto \frac{\kappa}{\Delta^2+\kappa^2}$ (indicated by the dashed line). (b) The ratio of oscillation period $T_\mathrm{BMF}/T_\mathrm{MF}$. The ratio is independent of dissipation $\kappa$. The parameters are fixed to $\Delta = 2\pi\times 40~\mathrm{MHz}$, $\eta=2\pi\times 1~\mathrm{kHz}$.}
    \label{MFvsQF}
\end{figure}

\begin{figure*}[!t]
    \centering
    \includegraphics[width=1\textwidth]{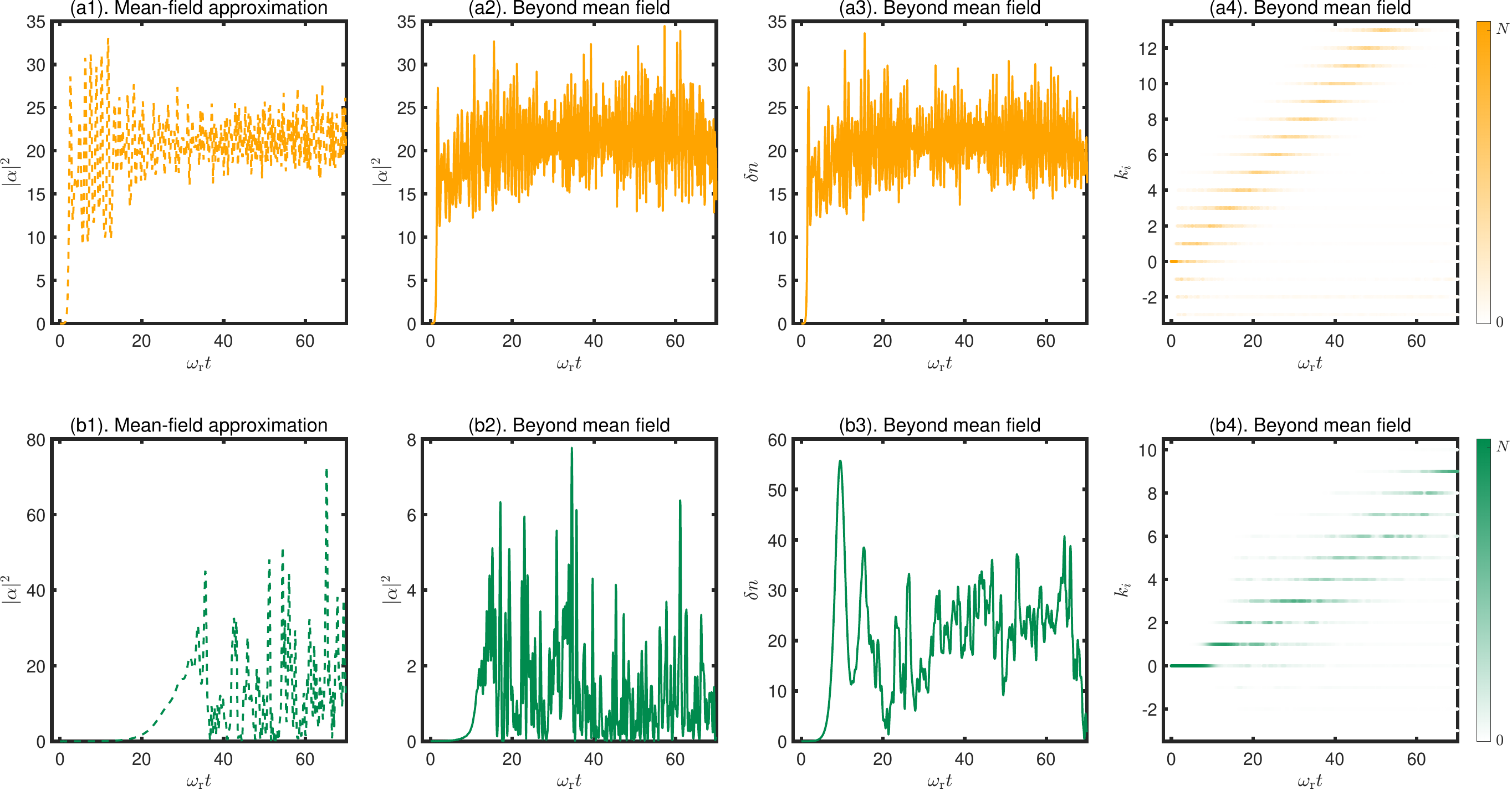}
    \caption{Evolution of the other two phases. (a1) The photon occupation in mean-field approximation $|\alpha|^2$. The parameters are $\Delta = -2\pi\times 30~\mathrm{MHz}$, $\eta=2\pi\times 2~\mathrm{kHz}$. (a2,3,4) Two part of the photon occupation, $|\alpha|^2$ and $\delta n$, and the atomic momentum distribution over time beyond mean-field approximation. (b). The lower panel is the same but in the dephasing phase, the parameters are $\Delta = 2\pi\times 10~\mathrm{MHz}$, $\eta=2\pi\times 1.5~\mathrm{kHz}$.}
    \label{Other}
\end{figure*}

Then, we also calculate the ratios between the oscillation period beyond and within mean-field approximation under different parameters. We found the ratios show a universality with the dissipation (see Fig.\ref{MFvsQF}). That is because the transfer rate depends on the dissipation in the same way. In our result, after a single oscillation, the correlations between different momenta all vanish (i.e. off-diagonal elements of density matrix $\langle \rho_{m,n \neq m} \rangle=0$). So the density matrix only contains uncorrelated momentum distribution, $\rho_{mn}=\rho_{nn}\delta_{mn}=f_n\delta_{mn}$. In the transfer cycle: $i\to i+1$, the evolution of $f_i=\rho_{ii}$ is 
\begin{eqnarray}
	\partial_t f_{i}(t)=-\frac{2\kappa\eta^2}{\Delta^2+\kappa^2}(f_{i}f_{i+1}+f_i-\delta n f_{i+1}).
\end{eqnarray}
While in the mean-field approximation, after eliminating $\alpha$, we obtain the coherent evolution of $f_i$:
\begin{eqnarray}
	\partial_t f_{i}(t)=-\frac{2\kappa\eta^2}{\Delta^2+\kappa^2}f_{i}f_{i+1}.
\end{eqnarray}
We can see the prefactor and the leading term are the same in both equations. This is non-trivial, since the former prefactor totally comes from the fluctuation and the latter prefactor totally comes from the coherent mean field. The ratio of transfer rate then only depends on the effect of subleading nonhomogeneous terms induced by fluctuation.

It should be careful to test our results in experiments. In typical cavity experiments, we generally measure how many photons leak out of the cavity to non-destructively deduce the photon occupation inside the cavity. In this sense,  while only measuring $\left\langle\hat{a}^\dag\hat{a}\right\rangle$, our results are qualitatively  similar to the original mean-field result, for they both give the phenomenon of periodic oscillation of the photon number. But quantitative measurement of the period can distinguish the difference. Or if the experiment can offer to observe the phase of leaking photon by a heterodyne setup, we can see its phase will be out of order.

Apart from the spatial-temporal lattice, the evolution under quantum fluctuation in the other two phases is also shown in Fig.\ref{Other}. It can be seen that their qualitative properties do not change. The lattice depth $|\alpha|$ in the spatial-lattice phase keeps almost constant and the evolution in the dephasing phase is still chaotic. However, quantatively, due to the fluctuation, the atomic transfer rate between momenta is enhanced by fluctuation in both phases.

\section{Summary}
In summary, we have investigated the behavior of one-dimensional bosons loaded in an optical cavity transversely pumped by a travelling wave beyond mean-field approximation. By including the fluctuation of the cavity field, we obtain a more precise set of equations of motion. Numerical results show that the spatial-temporal lattice does not exist in the mean-field level any more but survives in the fluctuation. From an experimental perspective, the expected oscillation of cavity photon occupation can still be observed, qualitatively like the previous conclusion. But quantitatively, the oscillating frequency will be enhanced. Besides, the other two phases are not significantly influenced.


This research brings a concrete reminder for us that the mean-field dynamics can be impacted drastically in non-equilibrium systems. Quantum fluctuation plays a significant role in the evolution. Future mean-field research on other novel non-equilibrium dynamical phenomena should be treated cautiously.

\section{ACKNOWLEDGMENTS}
\textit{Acknowledgment} We thank Yu Chen for helpful discussion. This work is supported by NSFC (Grant No. GG2030007011 and No. GG2030040453) and Innovation Program for Quantum Science and Technology (No.2021ZD0302004).

\bibliographystyle{unsrt}

\end{document}